\documentclass[reprint,
superscriptaddress,
amsmath,amssymb,
aps,
prl,
longbibliography,
showpacs
]{revtex4-1}
\usepackage{graphicx}
\usepackage{epstopdf}

\usepackage{color}

\begin{document}

\title{Unconditional entanglement interface for quantum networks}
\author{Christoph~Baune}
\affiliation{Institut f\"ur Laserphysik und Zentrum f\"ur Optische Quantentechnologien, Universit\"at Hamburg, Luruper Chaussee 149, 22761 Hamburg, Germany}
\affiliation{Institut f\"ur Gravitationsphysik, Leibniz Universit\"at Hannover and Max-Planck-Institut f\"ur Gravitationsphysik (Albert-Einstein-Institut), Callinstrasse 38, 30167 Hannover, Germany}
\author{Jan~Gniesmer}
\affiliation{Institut f\"ur Gravitationsphysik, Leibniz Universit\"at Hannover and Max-Planck-Institut f\"ur Gravitationsphysik (Albert-Einstein-Institut), Callinstrasse 38, 30167 Hannover, Germany}
\affiliation{Institut f\"ur Festk\"orperphysik, Leibniz Universit\"at Hannover, Appelstrasse 2, 30167 Hannover, Germany}
\author{Sacha~Kocsis}
\affiliation{Institut f\"ur Laserphysik und Zentrum f\"ur Optische Quantentechnologien, Universit\"at Hamburg, Luruper Chaussee 149, 22761 Hamburg, Germany}
\affiliation{Institut f\"ur Gravitationsphysik, Leibniz Universit\"at Hannover and Max-Planck-Institut f\"ur Gravitationsphysik (Albert-Einstein-Institut), Callinstrasse 38, 30167 Hannover, Germany}
\affiliation{Centre for Quantum Dynamics and Centre for Quantum Computation and Communication Technology, Griffith University, Brisbane 4111, Australia}\author{Christina~E.~Vollmer}
\author{Petrissa~Zell}
\affiliation{Institut f\"ur Gravitationsphysik, Leibniz Universit\"at Hannover and Max-Planck-Institut f\"ur Gravitationsphysik (Albert-Einstein-Institut), Callinstrasse 38, 30167 Hannover, Germany}
\author{Jarom\'ir~Fiur\'a\v{s}ek}
\affiliation{Department of Optics, Palack\'y University, 17. listopadu 12, 77146 Olomouc, Czech Republic}
\author{Roman~Schnabel}
\email{roman.schnabel@physnet.uni-hamburg.de}
\affiliation{Institut f\"ur Laserphysik und Zentrum f\"ur Optische Quantentechnologien, Universit\"at Hamburg, Luruper Chaussee 149, 22761 Hamburg, Germany}
\affiliation{Institut f\"ur Gravitationsphysik, Leibniz Universit\"at Hannover and Max-Planck-Institut f\"ur Gravitationsphysik (Albert-Einstein-Institut), Callinstrasse 38, 30167 Hannover, Germany}

\date{\today}

\begin{abstract}
Entanglement drives nearly all proposed quantum information technologies. 
By up-converting part of a 1550 nm two-mode squeezed vacuum state to 532 nm, we demonstrate the generation of strong continuous-variable entanglement between widely separated frequencies. 
Nonclassical correlations were observed in joint quadrature measurements of the 1550~nm and 532~nm fields, showing a maximum noise suppression 5.5 dB below vacuum. 
Our versatile technique combines strong nonclassical correlations, large bandwidth and, in principle, the ability to entangle the telecommunication wavelength of 1550~nm with any optical wavelength.
\end{abstract}

\maketitle
There has been continuous interest in investigating long-range quantum information networks, for reasons both fundamental~\cite{Salart2008a} and applied~\cite{Kimble2008}. 
Due to extremely long coherence times, and the relative ease with which they can be manipulated, optical modes are commonly considered to be the best means of distributing entanglement over large distances. 
Established fiber-optic technology offers relatively high transmission efficiency for wavelengths in the telecommunication band around 1550~nm, where scattering and absorption rates leading to photon loss are minimal. 

Entangling frequency modes of optical fields has attracted increased attention in recent years, as a quantum network would rely on interfacing light at telecommunication wavelengths with matter-based quantum memories that are addressable at visible wavelengths.
A true quantum information network would incorporate a number of nodes, where quantum states could be stored and even processed. 
These nodes could generally take the form of single atoms~\cite{Ritter2012}, atomic ensembles~\cite{Jensen2011,Reim2011} or solid-state systems~\cite{Zhong2015, Bernien2013}. 
This implies that the reversible mapping of quantum states between optical and material modes will necessarily be ubiquitous in quantum networks. 
As almost all current quantum memories operate at visible wavelengths, an efficient interface between entangled optical modes at telecommunication and visible wavelengths is a key capability to demonstrate. 

\begin{figure}[!b!]
	\centering
		\includegraphics[width=0.48\textwidth]{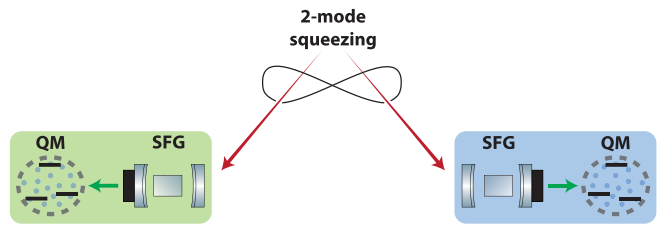}
	\caption{(Color online) Conceptual schematic of an elementary segment of a quantum network where entangled light beams establish quantum correlations between two nodes of the network. 
	Efficient transmission of light is ensured by operating at telecommunication wavelength, and the coupling to local quantum memories (QM) is enabled by frequency up-conversion (SFG) of the transmitted light beams.}
	\label{fig:EntDist}
\end{figure}

A conceptual scheme of an elementary quantum link between two nodes of a quantum network is illustrated in~Fig.~\ref{fig:EntDist}. 
A source emits two optical beams at telecommunication wavelength prepared in an entangled two-mode squeezed vacuum state, and each beam is sent to one node of the quantum network. 
To enable efficient interfacing with quantum memories, the transmitted light beams are frequency up-converted, and their quantum state is stored in quantum memories \cite{Lvovsky2009} for further processing. 
This schematic can represent, e.g., an elementary segment of a quantum repeater \cite{Briegel1998,Duan2001,Brask2010}, where efficient transmission of light over short distances would be combined with local processing of stored quantum states, entanglement distillation and entanglement swapping, to efficiently establish quantum correlations over arbitrarily large distances.

Previous work has demonstrated the interface between telecommunication and visible wavelengths with entangled photon pairs \cite{Tanzilli2005}, and storage in a quantum memory of single photons up-converted from telecom to visible wavelength has been reported \cite{Maring2014}.
In contrast to this discrete-variable (DV) encoding of quantum information, we focus on an interface for continuous-variables (CV), which is more general, as the states are not restricted to a limited Hilbert space. 
The main advantage of CV encoding is that it allows for deterministic quantum operations \cite{Furusawa1998, Marek2011, Takeda2013}, and benefits from well-established measurement techniques such as balanced homodyne detection. 
 There have been several demonstrations of quantum memories for CV states \cite{Julsgaard2004,Appel2008,Honda2008,Hosseini2011,Hedges2010,Jensen2011} and the classical benchmark for memory fidelity has been experimentally surpassed using displaced two--mode squeezed states \cite{Jensen2011}.

Here we deal with a broadband two-mode squeezed vacuum state which represents a fundamental resource for CV quantum information processing \cite{Braunstein2005}.  
We demonstrate and characterize frequency up-conversion of one half of this state from 1550~nm to 532~nm, and we verify that the quantum conversion preserves the modes' broadband entanglement.
It has been previously demonstrated that CV entanglement can span optical frequencies one octave apart \cite{Grosse2008}. 
However, no ``on the fly'' wavelength conversion has yet been achieved.
States with nonclassical characteristics over a broad frequency range are particularly interesting for channel multiplexing to achieve higher density information encoding \cite{Hage2010}, as well as for time-frequency domain CV quantum computing protocols \cite{Humphreys2014}. 
Our modular ``on the fly'' approach is versatile and can be modified to entangle 1550~nm light not only with 532~nm but with any other optical wavelength.

\begin{figure}[htb]
	\centering
		\includegraphics[width=0.45\textwidth]{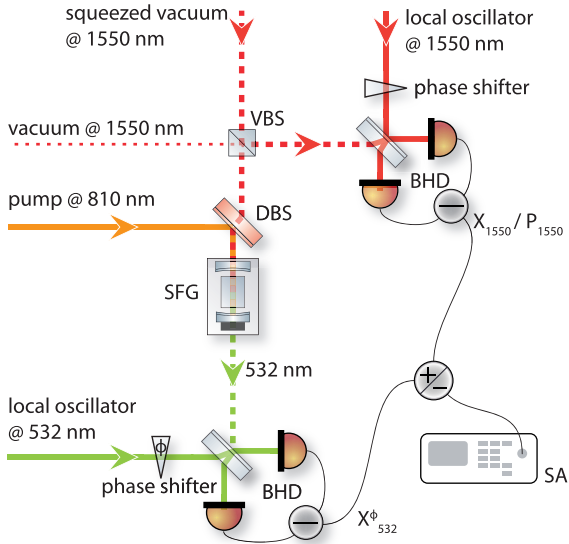}
	\caption{(Color online) Illustration of the experimental setup. 
	Squeezed vacuum states of continuous-wave light at 1550~nm are produced in a standing-wave cavity by degenerate type I optical parametric down-conversion, as described in more detail in Ref.~\cite{Baune2015}, and split up at a variable beam splitter. 
	The reflected part is directly sent to a balanced homodyne detection (BHD) while the reflected mode is up-converted to 532~nm and also detected in a BHD.
	The sum of the BHD signals is recorded by a spectrum analyzer (SA).}
	\label{fig:Setup}
\end{figure}
Our proof-of-principle demonstration of a versatile continuous variable entanglement interface is based on the efficient frequency up-conversion of one part of a broadband two-mode squeezed vacuum state of light.
The schematic of the experimental setup can be found in Fig.~\ref{fig:Setup}.
First, squeezed vacuum states of light at 1550~nm are produced by cavity-enhanced degenerate type I optical parametric down-conversion, also called optical parametric amplification (OPA).
The squeezing bandwidth is determined by the cavity bandwidth.
The squeezed states are subsequently mixed with the vacuum mode at a variable beam splitter (VBS), to generate entanglement between the reflected and transmitted mode \cite{Eberle2011}.
The transmitted mode is mode-matched to the sum-frequency generation cavity (SFG), which is pumped with a strong coherent field at 810~nm.
In the SFG the 1550~nm signal field is up-converted to 532~nm, while its quantum properties are maintained \cite{Kumar1990,Vollmer2014}.
The up-converted mode is analyzed at a balanced homodyne detector (BHD) utilizing a local oscillator at 532~nm.
A local oscillator at 1550~nm is used to analyze the reflected mode of the VBS at a second BHD.
Further details about the various components of the setup can be found in \cite{Samblowski2014,Baune2014,Baune2015}.

The entanglement of the two light beams at 532~nm and 1550~nm can be conveniently characterized by the quantity $\mathcal{I}$, introduced by Duan et al. \cite{Duan2000},
\begin{equation}
\mathcal{I}=\mathrm{Var}[\hat{X}_{1550}+\hat{X}_{532}]+\mathrm{Var}[\hat{P}_{1550}-\hat{P}_{532}],
\label{DuanI}
\end{equation} 
where $\hat{X}_\lambda$ and $\hat{P}_{\lambda}$ denote the amplitude and phase quadratures, respectively, of an optical beam at wavelength $\lambda$.
The quadrature variances are normalized such that $\mathrm{Var}[\hat{X}]=\mathrm{Var}[\hat{P}]=1$ for vacuum, and $\mathcal{I}<4$ certifies the presence of entanglement of the two beams \cite{Duan2000}.
The variances of linear combinations of quadratures appearing in Eq.~(\ref{DuanI}) characterize correlations between individual modes of a two-mode quantum state.
A simple theoretical model yields
\begin{eqnarray}
\mathrm{Var}[\hat{X}_{1550}+\hat{X}_{532}] &=& 2 -(1-V_{-})(t \tau_{532}+r \tau_{1550})^2, \nonumber \\
\mathrm{Var}[\hat{P}_{1550}-\hat{P}_{532}] &=& 2 +(V_{+}-1)(t \tau_{532}-r \tau_{1550})^2. \nonumber \\
\label{Variances}
\end{eqnarray}
Here $V_{-}$ and $V_{+}$ denote the variances of squeezed and anti-squeezed quadratures of the input state to the VBS. The amplitude transmittance and reflectance of the VBS are denoted by $t$  and $r$, respectively. The effective overall amplitude transmittances of the two output modes of the VBS, including final detection efficiencies in each of the BHDs, are denoted as $\tau_{532}$ and $\tau_{1550}$. 

\begin{figure}[htb]
	\centering
		\includegraphics[width=0.48\textwidth]{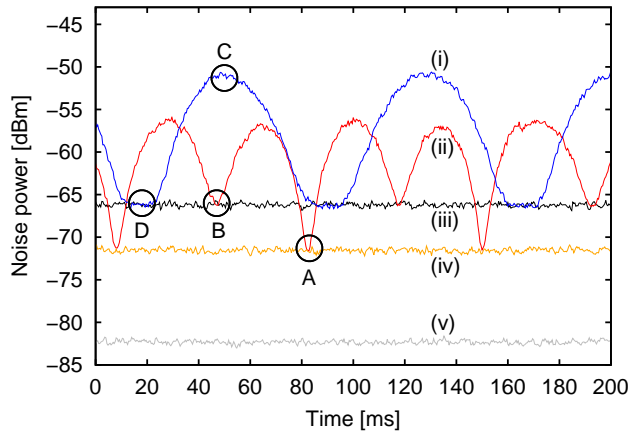}
	\caption{(Color online) Characterization of our unconditional quantum interface. Shown is the sum of the two BHD signals at 5~MHz sideband frequency. While the phase  $\phi$ of the BHD at 532~nm was continuously scanned, the phase of the BHD at 1550~nm was set to measure the squeezed quadrature amplitude $\hat{X}_{1550}$ (ii, red trace) or anti-squeezed $\hat{P}_{1550}$ quadrature (i, blue). The four extremal points represent the following measurement settings: A: $\rm{Var}[\hat{X}_{1550}+\hat{X}_{532}]$, B: $\mathrm{Var}[\hat{X}_{1550}-\hat{X}_{532}]$, C: $\mathrm{Var}[\hat{P}_{1550}+\hat{P}_{532}]$, D: $\mathrm{Var}[\hat{P}_{1550}-\hat{P}_{532}]$. The orange trace (iv) was recorded when the 532~nm phase was also fixed, revealing stable nonclassical correlations about 5.5~dB below the vacuum level (iii, black). Note that the traces were recorded successively, and there is no actual meaning in the relative positions of the minima and maxima. None of the traces was corrected for our detection scheme's dark noise (v, gray).}
	\label{fig:Ent}
\end{figure}

 In the experiment, the variable beam splitter is tuned such that the  anti-squeezed noise is fully canceled in the difference of the phase quadrature amplitudes $\hat{P}_{1550}-\hat{P}_{532}$. 
This is achieved when $t \tau_{532}=r\tau_{1550}$, and in this case the variance of the difference in phase quadratures reaches the vacuum noise level, $\mathrm{Var}[\hat{P}_{1550}-\hat{P}_{532}]=2$, see point D in Fig.~\ref{fig:Ent}. 
 Within the error bars of our experiment, this setting also provides the strongest Gaussian entanglement as quantified by $\mathcal{I}$. Indeed, it follows from Eq.~(\ref{Variances}) that, for this setting,  entanglement is certified and $\mathcal{I}<4$ whenever the input state is squeezed and $V_{-}<1$. 
 A more detailed theoretical analysis, however, reveals that the minimum value of $\mathcal{I}$ is attained at a marginally different beam splitter tuning \cite{Wagner2014}. In our experiment, the variable beam splitter sent a larger fraction of the squeezed input state to the entanglement interface to compensate for the non-perfect SFG conversion efficiency, as well as the non-perfect quantum efficiency of the photo diodes at 532~nm, both of which were approximately 90\%. In contrast, the mode at  1550~nm has less downstream losses, on the order of 12\%, dominated by the visibility of the balanced homodyne detector. 

The sum of the two BHD outputs is analyzed in a spectrum analyzer, providing noise powers of joint quadrature amplitudes that are proportional to the joint operator variances, over a range of sideband frequencies.
A zero span measurement of the revealed correlations at a sideband frequency of 5~MHz is shown in Fig.~\ref{fig:Ent}. 
While the phase of the 1550~nm BHD was set such that it measures the squeezed $\hat{X}_{1550}$ quadrature (on an auxiliary spectrum analyzer), the phase $\phi$ of the homodyne detector at 532~nm was scanned. 
The noise of the sum of the two detector signals is shown in red and strong correlations of about 5.5~dB below the vacuum noise are visible.
When the phase of the 1550~nm BHD is set to measure the anti-squeezed $\hat{P}_{1550}$ quadrature, the noise of the BHD sum only drops to the vacuum level.
The orange trace represents a measurement, where the quadrature at 1550~nm is squeezed and the 532~nm phase is set such that the minimum noise is measured, i.e. when $\phi=0$ so that $\hat{X}^\phi_{532}=\hat{X}_{532}$.
The measured correlations of $-5.5$~dB in the first measurement and $0$~dB in the second yield a Duan value  of $2.56$, which is significantly below the classical limit $4$. 

\begin{figure}[h]
	\centering
		\includegraphics[width=0.48\textwidth]{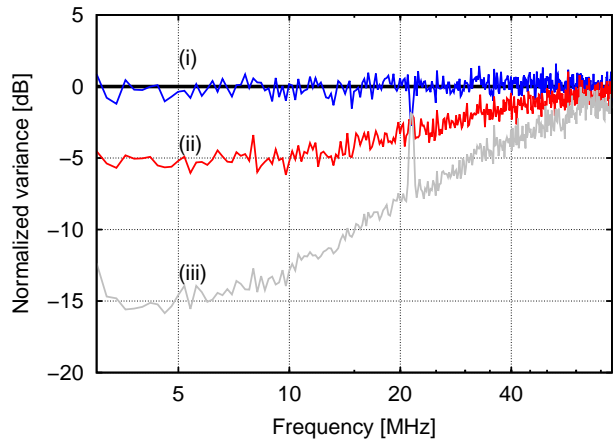}
	\caption{ (Color online) Spectral characterization of the entanglement interface. 
		Trace (i) shows the spectrum when $\mathrm{Var}[\hat{P}_{1550}-\hat{P}_{532}]$ is detected, corresponding to point D from Fig.~\ref{fig:Ent}. 
		Trace (ii) shows the spectrum for $\mathrm{Var}[\hat{X}_{1550}+\hat{X}_{532}]$, corresponding to point A from Fig.~\ref{fig:Ent}.  
		Normalized variance values below zero, the vacuum reference, signify nonclassical correlations.
		The bottom trace (iii) shows the dark noise of our balanced homodyne detector.}
	\label{fig:Spectrum}
\end{figure}

To analyze the spectral properties of the correlations, the frequency dependence of the sum signal was measured and is shown in Fig.~\ref{fig:Spectrum}. 
Therefore, the phase of the 1550~nm field was set to measure the squeezed (red trace) or anti-squeezed (blue) quadrature while the 532~nm phase was set to measure the minimum noise in the sum signal. 
More than 3~dB of correlations below the vacuum noise are measured up to a sideband frequency of about 20~MHz.

Entanglement was generated in our experiment by sending the squeezed vacuum state to a beam splitter, and mixing it with the vacuum state. 
The maximum strength of correlations between the output modes is limited in this method, but it was still sufficient for our proof-of-principle demonstration. 
Generating two-mode entanglement that is not limited in this way, in the original sense meant by Einstein, Podolsky and Rosen  \cite{Einstein1935}, only requires mixing two squeezed vacuum fields on a beam splitter. 
A fully controlled such source of EPR entanglement \cite{Reid1989} has previously been demonstrated by our group \cite{Eberle2013}.
Integrating it with the up-conversion set-up is technically more involved but straightforward in principle.

The storage of optical states in quantum memories is a key technology in setting up long-distance quantum networks \cite{Sangouard2011}. 
With minimal modifications, our current setup could be easily matched to any frequency single-mode CV quantum memory having a bandwidth in the MHz regime \cite{Simon2010}. 
The optical bandwidth can be varied by changing the mirror reflectivities of the squeezed light source and up-conversion cavities.
The optical frequency of the up-converted states can easily be changed by varying the frequency of the pump field. 
These capabilities are necessary to achieve high efficiency when integrating an optical entanglement source with a quantum memory.

Multi-mode quantum memories are also being investigated, particularly to achieve reasonable long-distance data transfer rates when they are used as quantum repeaters \cite{Afzelius2009,Sinclair2014}. 
Frequency multi-mode operation has recently been demonstrated with quantum memories based on atomic frequency combs (AFC) in rare-earth ion doped materials, and in principle these quantum memories could have a total accessible bandwidth in the range of hundreds of GHz \cite{Sinclair2014,Saglamyurek2015}. 
The total bandwidth of the states used in our approach can be multiplexed \cite{Hage2010} to take advantage of a large section of a multi-mode storage capacity.
In this sense a larger bandwidth implies a larger frequency space in which to encode distinct modes, and a nonclassical noise suppression of more than 3~dB over a bandwidth exceeding a GHz has been achieved at 1550~nm \cite{Ast2013}. 

Another promising application that could take advantage of our interface is for continuous-variable quantum computing in atomic memories using time-frequency entangled quantum modes \cite{Humphreys2014}. 
The proposal is based on the cluster-state quantum computing approach, which requires a large resource state with multi-mode entanglement.  
A squeezed vacuum field such as the one we characterized has a broad sideband spectrum that could be used for encoding these time-frequency quantum modes.

In conclusion, we present an entanglement-preserving interface for multi-color quantum optical networks. 
Our result demonstrates that current techniques in nonlinear optics enable the efficient frequency conversion of half of an entangled two-mode state. 
As a proof-of-principle, we up-converted from the near-infrared to the green spectrum, producing CV entanglement between continuous-wave light fields at 1550~nm and 532~nm. 
Up to 5.5~dB of nonclassical correlations with a bandwidth of about 20~MHz were maintained during the conversion process. 
The bandwidth can be increased by widening the cavity linewidths, and the strength of the nonclassical correlations can be further improved by reducing optical loss due to imperfect PIN photo diodes, cavity mode matchings and absorption in nonlinear crystals. 
The demonstrated entanglement interface in combination with quantum memories operating at visible wavelengths thus represents a promising and flexible alternative to direct quantum storage of telecom photons \cite{Jin2015,Saglamyurek2015}.
We consider our work to be an important building block for future multi-color quantum networks, which may further include CV as well as DV concepts. 

We thank Geoff Pryde and Vitus H\"andchen for useful discussions. 
This work was supported by the Deutsche Forschungsgemeinschaft (DFG), Germany, Project No. SCHN 757/4-1, by the Centre for Quantum Engineering
and Space-Time Research (QUEST), Germany and by the International Max Planck Research School for Gravitational Wave Astronomy (IMPRS-GW), Germany. 
This work was funded in part by the Australian Research Council (ARC) Centre of Excellence for Quantum Computation and Communication Technology, Australia (Project No. CE110001027). 
J.F. acknowledges financial support from the European Union (EU), Seventh Framework Program for Research (FP7), Project No. 308803, project name BRISQ2 cofinanced by Ministry of education, youth and sports of the Czech Republic (MSMT CR), Czech Republic, Project No. 7E13032. 
S.K. acknowledges financial support from the Alexander von Humboldt Foundation, Germany.

\end{document}